\documentclass[lettersize,journal]{IEEEtran}
\usepackage{amsmath,amsfonts}
\usepackage{algorithmic}
\usepackage{algorithm}
\usepackage{array}
\usepackage[caption=false,font=normalsize,labelfont=sf,textfont=sf]{subfig}
\usepackage{textcomp}
\usepackage{stfloats}
\usepackage{url}
\usepackage{verbatim}
\usepackage{graphicx}
\usepackage{cite}
\usepackage{algorithm}
\usepackage{algorithmic}
\usepackage{amssymb}
\usepackage{flushend}

\usepackage[colorlinks,
linkcolor=black,
anchorcolor=green,
citecolor=blue
]{hyperref}

\begin{document}

\title{Passive Beamforming For Practical RIS-Assisted Communication Systems With Non-Ideal Hardware}

\author{
	Yiming~Liu,
	Rui~Wang,~\IEEEmembership{Senior Member,~IEEE,}
	and
	Zhu~Han,~\IEEEmembership{Fellow,~IEEE}
\thanks{
	Yiming~Liu was with The Department of Information and Communication Engineering, Tongji University, Shanghai 201804, China. 
	He is now with The Edward S. Rogers Sr. Department of Electrical and Computer Engineering, University of Toronto, Toronto, Ontario M5S 3G4,
	Canada (e-mail: eceym.liu@mail.utoronto.ca).
	
	Rui~Wang is with The Department of Information and Communication Engineering, Tongji University, Shanghai 201804, China (e-mail: ruiwang@tongji.edu.cn).
	
	Zhu~Han is with The Department of Electrical and Computer Engineering,
	University of Houston, Houston, Texas 77004 USA (e-mail: zhan2@uh.edu).
	 }
}




\maketitle

\begin{abstract}
Reconfigurable intelligent surface (RIS) technology is a promising solution to improve the performance of existing wireless communications. To achieve its cost-effectiveness advan-tage, there inevitably exist certain hardware impairments in the system. Therefore, it is more reasonable to design passive beamforming in this scenario. Some existing research has considered such problems under transceiver impairments. However, their performance still leaves room for improvement, possibly due to their algorithms not properly handling the fractional structure of the objective function. To address this, the passive beamforming is redesigned in this correspondence paper, taking into account both transceiver impairments and the practical phase-shift model. We tackle the fractional structure of the problem by employing the quadratic transform. The remaining sub-problems are addressed utilizing the penalty-based method and the difference-of-convex programming. Since we provide closed-form solutions for all sub-problems, our algorithm is highly efficient. The simulation results demonstrate the superiority of our proposed algorithm. 
\end{abstract}

\begin{IEEEkeywords}
Reconfigurable intelligent surface (RIS), passive beamforming, fractional programming, difference-of-convex programming, hardware impairments, practical phase shifts. 
\end{IEEEkeywords}

\section{Introduction}
\IEEEPARstart{T}{he} explosive growth of data traffic requires to enhance the performance of existing wireless communications. To this end, a variety of wireless technological advances, including millimeter-wave (mmWave) communication and massive multiple-input multiple-output (MIMO), have been proposed. However, these technologies involve expensive hardware and consume immoderate energy. Hopefully, with the development of electromagnetic materials, reconfigurable intelligent surface (RIS) has emerged as a cost-effective promising technology to provide significant performance gain for wireless communications \cite{8741198, 9110869, 9326394, 8811733}.

Accordingly, RIS-assisted wireless communication systems have drawn significant research interest, especially the transmission optimization for various RIS-assisted wireless communication systems \cite{8811733, 8982186, 8930608, 9133130, 8941080, 8743496}. Specifically, transmit power minimization under the quality-of-service (QoS) constraints were investigated for RIS-assisted MISO systems \cite{8811733, 8930608}.
Some other objectives of optimization, such as achievable rate maximization \cite{8982186} and energy efficiency maximization \cite{8741198}, have also been considered carefully. 
Further, from the security perspective, the achievable secrecy rates have been maximized in \cite{9133130, 8743496}.
Another interesting perspective is energy harvesting. Authors of \cite{8941080} focus on an RIS-aided multi-antenna simultaneous wireless information and power transfer system. The purpose is to maximize the weighted sum power received by energy harvesting receivers while fulfilling the QoS constraints for information decoding receivers.
As shown by all the above works, the redesign of beamforming for RIS-assisted systems is important and inevitable. This is attributed to the fact that the phase shifts of an RIS are also coupled with the transmit beamforming.
Another important reason is that, in order to harness the benefits introduced by RIS, new constant-modulus constraints render more challenging problems to address.
It is worth noting that the existing works are predominantly conducted upon the assumption of ideal hardware. However, the central advantage of RIS-assisted wireless systems relies on the cost-effectiveness; that is to say, in order to achieve this, the system's hardware may inevitably incur some impairments, which may lead to severe performance degradation.
Therefore, it is necessary to redesign the passive beamforming for a non-ideal and more practical scenario.

Some existing works have already considered this problem.
However, due to the complicated structure of the optimization problem when consider hardware impairments, many existing works utilize the gradient ascent or descent method to address it, such as \cite{2006.00664, 9534477, 9570811}.
The minorization-maximization (MM)-based algorithm is another commonly used method \cite{9239335} to solve this problem.
These methods cannot utilize the fractional structure of the optimization problem and the performance is not good enough, especially when the signal-to-noise (SNR) is low, the pathloss is large, or the number of antennas at the base station is small.
Moreover, these methods cannot be extended to the case where the RIS has a practical phase shifts.

In this paper, we intend to maximize the spectrum efficiency for the RIS-assisted wireless communication system with both transceiver hardware impairments and practical phase shifts. We first transform the problem into a more intuitive form, and then we utilize its fractional structure to handle it. Specifically, we utilize the quadratic transform, the penalty-based method, and the difference-of-convex (DC) programming to address the problem.
We provide the closed-form solutions for all the sub-problems.
Simulation results demonstrate the superiority of our proposed algorithm compared to existing methods.

\section{System Model}

\begin{figure}[!t]
	\centering
	\includegraphics[width = 2.75 in]{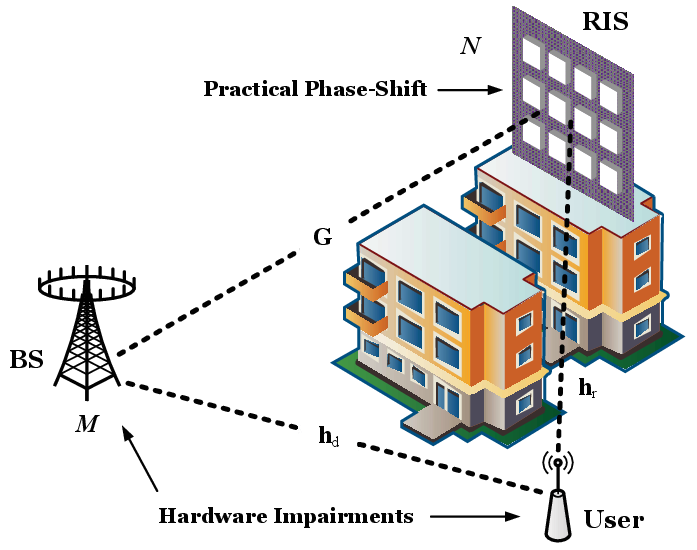}
	\caption{An RIS-assisted system with an $M$-antenna BS, a single-antenna user, and an RIS comprising $N$ reflecting elements.}
	\vspace{- 5 pt}
	\label{sysmod}
\end{figure}

In this paper, we consider an RIS-assisted system as shown in Fig. \ref{sysmod}, in which an RIS is deployed to enhance the wireless communication link between a multi-antenna base station (BS) and a single-antenna user.
The number of antennas at the BS is denoted as $M$, and the number of reflecting elements at the RIS is denoted as $N$.

\subsection{Channel Model}

The channels are assumed to have both line-of-sight (LOS) and non-line-of-sight (NLOS) components \cite{9181610, 9371709, 9312125}.
The direct channel $\mathbf{h}_{\mathrm{d}}$ from the base station to the user is given by
\begin{equation}
	\mathbf{h}_{\mathrm{d}}
	= \sqrt{ \frac{ \nu_{\mathrm{d}} {}_{\mathop{}} \omega_{\mathrm{d}} }{ \omega_{\mathrm{d}} + 1 } } \mathbf{\bar{h}}_{\mathrm{d}}
	+ \sqrt{ \frac{ \nu_{\mathrm{d}} }{ \omega_{\mathrm{d}} + 1 } } \mathbf{\tilde{h}}_{\mathrm{d}} \in \mathbb{C}^{M \times 1} ,
\end{equation}
where $\nu_{\mathrm{d}}$ models the large scale fading, $\omega_{\mathrm{d}}$ is the Rician factor, $\mathbf{\bar{h}}_{\mathrm{d}}$ and $\mathbf{\tilde{h}}_{\mathrm{d}}$ are the LOS and NLOS components, respectively.
For the specific model of LOS and NLOS components, please refer to \cite{9181610, 9371709, 9312125, 9847080, 9500437}, which will not be repeated here.
In a similar manner, the reflecting channels, $\mathbf{G}$ from the BS to the RIS and $\mathbf{h}_{\mathrm{r}}$ from the RIS to the user, can also be accurately modeled.

\subsection{Practical Equipment Model}

In this section, we establish a non-ideal and more practical system model where the amplitudes of reflecting elements are phase-dependent, and both the transceivers at the BS and the user suffer from hardware impairments.
The operation matrix of the RIS is defined as
\begin{equation}
	\mathbf{\Phi} 
	= \mathop{\mathrm{diag}} \left(\left[  
		\alpha_1  \exp \left( j\theta_1 \right), 
		\cdots, 
		\alpha_N  \exp \left( j\theta_N \right) \right]^{\mathsf{T}}
		\right),
\end{equation}
where $\theta_n \in [-\pi, \pi)$ and $\alpha_n \in [0, 1]$ represent the phase shift and the corresponding amplitude, respectively. We consider the practical phase-shift model in \cite{9115725}, and the phase-dependent amplitude response $\alpha_n (\theta_n)$ is
\begin{equation}
	\alpha_n \left( \theta_n \right)
	= \left( 1 - \alpha_{\min} \right) \left( \frac{ \; \sin (\theta_n - \phi) + 1 \; }{2} \right)^{\gamma} + \alpha_{\min},
\end{equation}
where $\alpha_{\min} \geq 0$, $\phi \geq 0$, and $\gamma \geq 0$ are the constants related to the circuit implementation.

The hardware impairments cause the distortion between the intended signal and the actual signal at transceivers, which can be well-modeled as uncorrelated additive Gaussian noise \cite{6891254}.
We utilize $\boldsymbol{\eta}_{\mathrm{b}} \sim \mathcal{CN} (0, \mathbf{\Upsilon}_{\mathrm{B}})$ and $\eta_{\mathrm{u}} \sim \mathcal{CN} (0,  v_{\mathrm{u}}^2)$ denote the distortion noise at the BS and the user, respectively.
Because the distortion noise power at an antenna is proportional to the signal power at this antenna \cite{9534477, 9239335, 6891254, 9650619}, we have
\begin{subequations}
	\begin{align}
		\mathbf{\Upsilon}_{\mathrm{B}}
		& = \rho_{\mathrm{b}} \mathop{\mathrm{diag}} \left( \mathbb{E} \left\lbrace \mathbf{s} \mathbf{s}^{\mathsf{H}} \right\rbrace \right), \\
		v_{\mathrm{u}}^2 \;
		& = \rho_{\mathrm{u}} \mathbf{h}^{\mathsf{H}\;} \mathbb{E} \left\lbrace ( \mathbf{s} + \boldsymbol{\eta}_{\mathrm{b}} ) ( \mathbf{s} + \boldsymbol{\eta}_{\mathrm{b}})^{\mathsf{H}} \right\rbrace \mathbf{h},
	\end{align}
\end{subequations}
where $\rho_{\mathrm{b}}$ and $\rho_{\mathrm{u}}$ are the proportionality coefficients which characterize the levels of hardware impairments at the BS and the user, respectively, and $\mathbf{h} = \mathbf{h}_{\mathrm{d}} + \mathbf{G} \mathbf{\Phi} \mathbf{h}_{\mathrm{r}}$ represents the overall channel.

\subsection{Received Signal Model}

According to the prior established model, the received signal at the user $y \in \mathbb{C}$ is given by
\begin{equation}
	y
	= ( \mathbf{h}_{\mathrm{d}} + \mathbf{G} \mathbf{\Phi} \mathbf{h}_{\mathrm{r}} )^{\mathsf{H}} ( \mathbf{s} + \boldsymbol{\eta}_{\mathrm{b}} ) + \eta_{\mathrm{u}} + n,
\end{equation}
where $\mathbf{s} \in \mathbb{C}^{M \times 1}$ is the transmitted signal, $n \in \mathbb{C}$ is the noise drawn from $\mathcal{CN} (0, \sigma_{\mathrm{u}}^2)$, and the power utilized to transmit the intended signal at the BS is $p_{\mathrm{b}} = \mathbb{E} \{ \mathbf{s} \mathbf{s}^{\mathsf{H}} \} $.
Then, the received signal-to-noise ratio (SNR) at the user is
\begin{equation} \label{SNR}
	\mathcal{S}
	= \frac{ \mathbf{h}^{\mathsf{H}} \mathbf{s} \mathbf{s}^{\mathsf{H}} \mathbf{h} }
	{ \rho_{\mathrm{b}} \mathbf{h}^{\mathsf{H}} \mathop{\mathrm{diag}} \left( \mathbf{s} \mathbf{s}^{\mathsf{H}} \right) \mathbf{h} + v_{\mathrm{u}}^2 + \sigma_{\mathrm{u}}^2 } .
\end{equation}

\vspace{5 pt}
\section{Beamforming Optimization}

In this section, we will propose an algorithm to optimize the passive beamforming in a non-ideal and practical RIS-assisted system. Our objective is to maximize the spectrum efficiency by adjusting the phase shifts of the RIS. Before achieving this purpose, we first formulate the optimization problem.

\subsection{Problem Formulation}

In our previous work \cite{9322510}, the maximum-ratio transmission (MRT) has been proven to be the optimal solution of transmit beamforming, even with non-ideal hardware. 
Then, (\ref{SNR}) can be further written as
\begin{equation}
	\mathcal{S}
	= \mathbf{h}^{\mathsf{H}} \left[ \rho \mathop{\mathrm{diag}} \left( \mathbf{h} \mathbf{h}^{\mathsf{H}} \right) + \rho_{\mathrm{u}} \mathbf{h} \mathbf{h}^{\mathsf{H}} + \sigma^2 \mathbf{I} \right]^{-1} \mathbf{h} ,
\end{equation}
where $\rho = \rho_{\mathrm{b}} ( 1 + \rho_{\mathrm{u}} )$ and $\sigma^2 = \frac{\sigma_{\mathrm{u}}^2}{p_{\mathrm{b}}}$.
By using (2.2) in \cite{SILVERSTEIN1995175}, the spectrum efficiency of downlink can be transformed as
\begin{equation} \label{SE}
	\mathcal{C}
	= \log_2 \left[1 + \frac{\mathbf{h}^{\mathsf{H}} \left[ \rho \mathop{\mathrm{diag}} \left( \mathbf{h} \mathbf{h}^{\mathsf{H}} \right) + \sigma^2 \mathbf{I} \right]^{-1} \mathbf{h}}
	{ 1 + \rho_{\mathrm{u}} \mathbf{h}^{\mathsf{H}} \left[ \rho \mathop{\mathrm{diag}} \left( \mathbf{h} \mathbf{h}^{\mathsf{H}} \right) + \sigma^2 \mathbf{I} \right]^{-1} \mathbf{h} } \right] .
\end{equation}
This formula has a structure of $f(x) = \log_2 (1 + \frac{x}{1+ax})$, $a > 0$, which is a monotonically increasing function.
To maximize the spectrum efficiency, we only need to maximize the numerator.
In addition, we define a new matrix as $\boldsymbol{\mathbf{H}} = \mathbf{G} \mathop{\mathrm{diag}} (\mathbf{h}_{\mathrm{r}})$.
Then, the overall channel can be equivalently rewritten as $\mathbf{h} = \mathbf{h}_{\mathrm{d}} + \boldsymbol{\mathbf{H}} \mathop{\mathrm{arcdiag}} (\mathbf{\Phi})$.
Then, the objective function can be formulated as follows:
\begin{equation} \label{}
	\begin{aligned}
		\mathcal{Q}_1 (\mathbf{x})
		& = \mathbf{h}^{\mathsf{H}} \left[ \rho \mathop{\mathrm{diag}} \left( \mathbf{h} \mathbf{h}^{\mathsf{H}} \right) + \sigma^2 \mathbf{I} \right]^{-1} \mathbf{h} \\
		& = \sum_{m=1}^{M} \frac{ \left\|  \mathrm{h}_{\mathrm{d}, m} + \mathbf{v}_m^{\mathsf{T}} \mathbf{x} \right\|^2 }{ \rho \left\| \mathrm{h}_{\mathrm{d}, m} + \mathbf{v}_m^{\mathsf{T}} \mathbf{x} \right\|^2 + \sigma^{2} }  \\
		& \triangleq \sum_{m=1}^{M} \frac{ A_m^2 (\mathbf{x}) }{ \rho A^2_m (\mathbf{x}) + \sigma^{2} },
	\end{aligned}
\end{equation}
where $\mathbf{v}_m^{\mathsf{T}}$ is the $m$-th row vector of $\mathbf{H}$, and $\mathbf{x} = \mathop{\mathrm{arcdiag}} (\mathbf{\Phi})$ denotes the optimization variable, with each entry representing the adjusted status of the corresponding reflecting element.

Then, the optimization problem for passive beamforming is formulated as follows:
\begin{subequations}
\label{FP}
	\begin{align}
		(\text{P1}): \;
		\mathop{\mathrm{maximize}}_{\mathbf{x}} \label{fx}\;
		& \; \mathcal{Q}_1 (\mathbf{x}) \\
		\mathop{\mathrm{s.t.}} \;\;\;\;\;
		& \; \left| x_n \right| = \alpha_n ( \mathop{\mathrm{arg}} \left( x_n \right) ), {\quad} \forall n ,  \label{nonCvx}.
	\end{align}
\end{subequations}
Problem (P1) is a class of fractional programming (FP) since $\mathcal{Q}_1 (\mathbf{x})$ has a sum-of-ratio structure, and all the numerators and denominators contain optimization variables, making (P1) difficult to solve.
Another challenge is dealing with the non-convex constraints in (\ref{nonCvx}).

\subsection{Quadratic Transform for (P1)}

The conventional algorithms for FP, such as Charnes-Cooper Transform \cite{charnes1962programming} and Dinkelbach's Transform \cite{dinkelbach1967nonlinear}, can perform very well for single-ratio cases but not for multiple-ratio cases.
The main reason is because the optimal values of the objective functions transformed by these algorithms are not necessarily the same as the values of the original objective functions. 
Therefore, when multiple ratios are involved, these algorithms cannot be applied to each ratio individually.

To further deal with the multi-ratio structure of the objective function in (P1), we employ the Quadratic Transform proposed in \cite{8314727}.
Then, problem (P1) is equivalent to
\begin{subequations} \label{transf}
	\begin{align}
	\mathop{\mathrm{maximize}}_{\mathbf{x}, \left\lbrace \lambda_m \right\rbrace} \;
		& \sum_{m=1}^{M} 2 \lambda_m A_m (\mathbf{x}) - \lambda_m^2 \left( \rho A_m^{2} (\mathbf{x}) + \sigma^{2} \right) \label{newob} \\
		\mathop{\mathrm{s.t.}} \;\;\;\;\;
		& {}_{\mathop{}} \left| x_n \right| = \alpha_n ( \mathop{\mathrm{arg}} \left( x_n \right) ), {\quad} \forall n , \\
		& \; \lambda_m \in \mathbb{R}, {\;\;\;\;\;\;\;\;\;\;\;\;\;\;\;\;\;\;\;\;\;\;}  \forall m, 
	\end{align}
\end{subequations}
where $\lambda_m, \forall m$ are the new auxiliary variables. 
By introducing $\lambda_m, \forall m$, the numerators and denominators can be decoupled, and the new problem has the equivalent optimal solution with the original problem. Most importantly, (\ref{newob}) is equivalent to $\mathcal{Q}_1 (\mathbf{x})$, i.e., let $\boldsymbol{\lambda} = \arg \max_{\boldsymbol{\lambda}} (\text{\ref{newob}})$ for certain $\mathbf{x}$, and then (\ref{newob}) is equivalent to the original objective function for this $\mathbf{x}$. Another good property of this transform is that (\ref{newob}) is concave over $\boldsymbol{\lambda}$ for fixed $\mathbf{x}$.

Next, we solve the new problem transformed in (\ref{transf}). 
When $\mathbf{x}$ is held fixed, (\ref{newob}) is concave over $\boldsymbol{\lambda}$. Then, the optimal $\boldsymbol{\lambda}^{\star}$ can be easily found in closed form as
\begin{equation} \label{lambda}
	\lambda_m^{\star}
	= \frac{ A_m (\mathbf{x}) }{ \rho A_m^2 (\mathbf{x}) + \sigma^{2} }, \quad \forall m.
\end{equation}
When $\lambda_m, \forall m$ are fixed, the optimization sub-problem for solving $\mathbf{x}$ is given by
\begin{subequations}
	\begin{align}
		(\text{P2}): \;
		\mathop{\mathrm{maximize}}_{\mathbf{x}} \;
		& \sum_{m=1}^{M} 
		 2 \lambda_m A_m (\mathbf{x}) - \lambda_m^2 \left( \rho A_m^2 (\mathbf{x}) + \sigma^{2} \right) \label{P3.2-OF} \\
		\mathop{\mathrm{s.t.}} \;\;\;\;\;
		& \; \left| x_n \right| = \alpha_n ( \mathop{\mathrm{arg}} \left( x_n \right) ), {\quad} \forall n . \label{P3.2-C1}
	\end{align}
\end{subequations}
Problem (P2) is still difficult to handle. In the sequel, we will deal with the non-convex objective function and constraints.

\subsection{Penalty Based Algorithm for (P2)}
We deal with the non-convex constraints in (\ref{P3.2-C1}) by adding a constraint-related penalty term to (\ref{P3.2-OF}). To this end, the new objective function is
\begin{equation}
	\begin{aligned}
		\mathcal{Q}_2 \left( \mathbf{x}, \left\lbrace \theta_n \right\rbrace \right) 
		= \sum_{m=1}^{M} \left[2 \lambda_m A_m (\mathbf{x}) - \lambda_m^2 \left( \rho A_m^2 (\mathbf{x}) + \sigma^{2} \right) \right] \; & \\
		- \mu \sum_{n = 1}^{N} \left| x_n - \alpha_n \left( \theta_n \right) e^{j \theta_n} \right|^2 & ,
	\end{aligned}
\end{equation}
Then, problem (P2) can be reformulated as	
\begin{subequations}
	\begin{align} 
		\mathop{\mathrm{maximize}}_{\mathbf{x}, \left\lbrace \theta_n \right\rbrace} \;
		& \; \mathcal{Q}_2 \left( \mathbf{x}, \left\lbrace \theta_n \right\rbrace \right) \\
		\mathop{\mathrm{s.t.}} \quad\;\; 
		& - \pi \leq \theta_n \leq \pi, {\quad\;\;\;\;\;\;} \forall n .
	\end{align}
\end{subequations}
This problem can be solved in an iteration manner.
In the $t$-th round of iteration,
for given $\left\lbrace \theta_{n, t} \right\rbrace_{n=1}^{N}$,
we define a vector as
\begin{equation}
	\boldsymbol{\theta}_{t} 
	= \left[  
	\alpha_1 \left( \theta_{1, t} \right)  \exp \left( j \theta_{1, t} \right), \cdots, \alpha_N \left( \theta_{N, t} \right) \exp \left( j \theta_{N, t} \right) 
	\right]^{\mathsf{T}}.
\end{equation}
Then, $\mathbf{x}$ can be optimized by solving the following problem:
\begin{equation}
	\label{P3.1}
	\begin{aligned}
		(\text{P2.1}): \mathop{\mathrm{maximize}}_{\mathbf{x}} \;
		& ( \sum_{m=1}^{M} 2 \lambda_m A_m (\mathbf{x}) )  \\
		- & \mathop{} (\sum_{m=1}^{M} \lambda_m^2 \left( \rho A_m^2 (\mathbf{x}) + \sigma^{2} \right)
	    + \mu \left\| \mathbf{x} - \boldsymbol{\theta}_{t} \right\|^2) \\
		 \triangleq & \; f_1 (\mathbf{x}) - f_2 (\mathbf{x}) .
	\end{aligned}
\end{equation}
Since $f_1 (\mathbf{x})$ and $f_2 (\mathbf{x})$ are both convex, this problem is a class of DC programming.
We utilize the concave-convex procedure (CCCP) \cite{7547360, NIPS2001_a0128693, lipp2016variations} to solve it, i.e., at the $t$-th round of iteration, $\mathbf{x}_{t+1}$ is updated by solving the following subproblem:
\begin{equation} \label{CCCP}
	\mathop{\mathrm{minimize}}_{\mathbf{x}} \; f_2 (\mathbf{x}) - \left( f_1 (\mathbf{x}_{t}) + \nabla f_1^{\mathsf{T}} (\mathbf{x}_{t}) \left( \mathbf{x} - \mathbf{x}_{t} \right) \right) 
\end{equation}
where 
\begin{equation}
	\nabla f_1 (\mathbf{x}_{t})
	= \sum_{m=1}^{M} \frac{\lambda_m}{A_m (\mathbf{x}_{t})} \left( \mathrm{h}_{\mathrm{d}, m}^{*} \mathbf{v}_m + \mathbf{v}_m \mathbf{v}_m^{\mathsf{H}} \mathbf{x}_{t}^{*} \right).
\end{equation}
Then, the closed-form optimal solution can be achieved by setting the gradient of (\ref{CCCP}) with respect to $\mathbf{x}$ to zero. And for simplicity, we utilize the matrix representation to eliminate the summation symbols, which is given by
\begin{equation}
	\begin{aligned}
		\mathbf{x}_{t+1}^{*}
		& = \left( \mathbf{H}^{\mathsf{T}} \operatorname{diag} ( \rho \boldsymbol{\lambda} \boldsymbol{\lambda}^{\mathsf{T}} ) \mathbf{H}^{*} + \mu \mathbf{I} \right)^{-1} \\ 
		& \;\;\;\; \times \left( \nabla f_1 (\mathbf{x}_{t}) - \operatorname{diag} ( \rho \boldsymbol{\lambda} ) \mathbf{H}^{\mathsf{T}} \mathbf{h}_{\mathrm{d}}^{*} + \mu\boldsymbol{\theta}_{t}^{*} \right),
	\end{aligned} 
\end{equation}
where the $m$-th element of $\boldsymbol{\lambda}$ is $\lambda_m$, $\forall m$.
For any given $\mathbf{x}_{t+1}$, $\{ \theta_n \}_{n=1}^{N}$ can be optimized by
\begin{subequations}
	\begin{align}
		(\text{P2.2}): \; \mathop{\mathrm{maximize}}_{\left\lbrace \theta_n \right\rbrace} \;
		& - \sum_{n = 1}^{N} \left| x_n - \alpha_n \left( \theta_n \right) e^{j \theta_n} \right|^2 \\
		\mathop{\mathrm{s.t.}} \quad\;\; 
		& - \pi \leq \theta_n \leq \pi, {\quad\;\;\;\;\;\;\;} \forall n .
	\end{align}
\end{subequations}
A closed-form approximate solution can be obtained by using the trust-region method, similar to [\citenum{9115725}, {Proposition 3}]. 
Another efficient solution is the line search algorithm. It is sufficiently good in this problem since the optimal $\theta_n$ is slightly deviates away from $\mathop{\mathrm{arg}} (x_n)$.
The overall procedure of our proposed optimization algorithm to solve (P1) is shown in Algorithm 1.

\vspace{10 pt}

\begin{algorithm}[!h]
\vspace{3 pt}
\caption{The Proposed Algorithm for Solving (P1)}
\begin{algorithmic}[1]
	\STATE \textbf{input} The channel matrices $\mathbf{h}_{\mathrm{d}}$, $\mathbf{H}$, the constants $\alpha_{\min}$, $\phi$, $\gamma$, and the hardware impairment coefficients $\rho_{\mathrm{b}}$ and $\rho_{\mathrm{u}}$.
		
	\STATE Initialize $\mathbf{x}$ to a feasible value.	
	\STATE Reformulate problem (P1) by the quadratic transform.
		
	\REPEAT
		 \STATE Update $\lambda_m, \forall m$ by (\ref{lambda}).	
		 \STATE Update $\mathbf{x}$ by solving problem (P2) as follows:
		 \REPEAT
		 	\STATE Update $\mathbf{x}$ by solving problem (P2.1).	
		 	\STATE Update $ \theta_n, \forall n$ by solving problem (P2.2).	
		 \UNTIL{convergency}
	\UNTIL{convergency}
		
	\STATE \textbf{output} The optimized phase shifting vector $\mathbf{x}$.
	\vspace{1 pt}
\end{algorithmic}
\end{algorithm}

\vspace{-5 pt}
\section{Numerical Results}

In this section, we will demonstrate the numerical results of our solution. Before we compare our results with the results of some existing works. We first give the upper bound of spectral efficiency, which is also the upper bound on the optimal result of the optimization problem. However, it is a loose bound since it is derived asymptotically. The related theorems are derived in our prior work \cite{9322510}, which are described briefly below.

Due to the existence of hardware impairments, the spectrum efficiency of RIS-assisted communication systems is bounded. Here, we denote the two types of asymptotic channel capacities as $\mathop{}\mathcal{C}_{1}$ and $\mathop{}\mathcal{C}_{2}$ when the transmit power of the BS and the number of reflecting elements approach to infinity, respectively.
Then, it holds that \cite{9322510}
\begin{equation} \label{up}
	\mathcal{C}_{1} = \mathcal{C}_{2}
	\leq \log_2 \left( 1 + \frac{M}{\rho_{\mathrm{b}} + \rho_{\mathrm{u}} (M + \rho_{\mathrm{b}})} \right).
\end{equation}

Since the structure of the optimization problem is very complicated when consider hardware impairments, many existing works utilize the gradient ascent/descent method to solve, such as \cite{2006.00664, 9534477, 9570811}. Here, to deal with the modulus constraints in problem (P1), we utilize the projected gradient ascent algorithm proposed in \cite{9534477} as a benchmark. This algorithm projects the solution of each step onto the closest feasible point satisfying the modulus constraint. The derivative of $\mathcal{Q}_1 (\mathbf{x})$ with respect to $\mathbf{x}$ is given by
\begin{equation}
	\begin{aligned}
		\nabla_{\mathbf{x}} \mathcal{Q}_1 (\mathbf{x})
		= \sum_{m=1}^{M} 
		& \frac{\mathrm{h}_{\mathrm{d}, m}^{*} \mathbf{v}_m + \mathbf{v}_m \mathbf{v}_m^{\mathsf{H}} \mathbf{x}^{*}}{ \rho A^2_m (\mathbf{x}) + \sigma^{2} } \\
		& - \frac{ \rho \left( \mathrm{h}_{\mathrm{d}, m}^{*} \mathbf{v}_m + \mathbf{v}_m \mathbf{v}_m^{\mathsf{H}} \mathbf{x}^{*} \right) A^2_m (\mathbf{x})}{\left( \rho A^2_m (\mathbf{x}) + \sigma^{2} \right)^2}.
	\end{aligned}
\end{equation}
In addition, we utilize the minorization-maximization (MM)-based method proposed in \cite{9239335} as another benchmark, and the update rule in each round of iterations is given by (15) in \cite{9239335}.

Overall, we compare the results of our proposed algorithm with these existing methods as well as the loose upper bound in (\ref{up}) and the randomized phase-shifting solution.

\begin{figure}[h]
	\centering
	\vspace{-10 pt}
	\includegraphics[width=3.2 in]{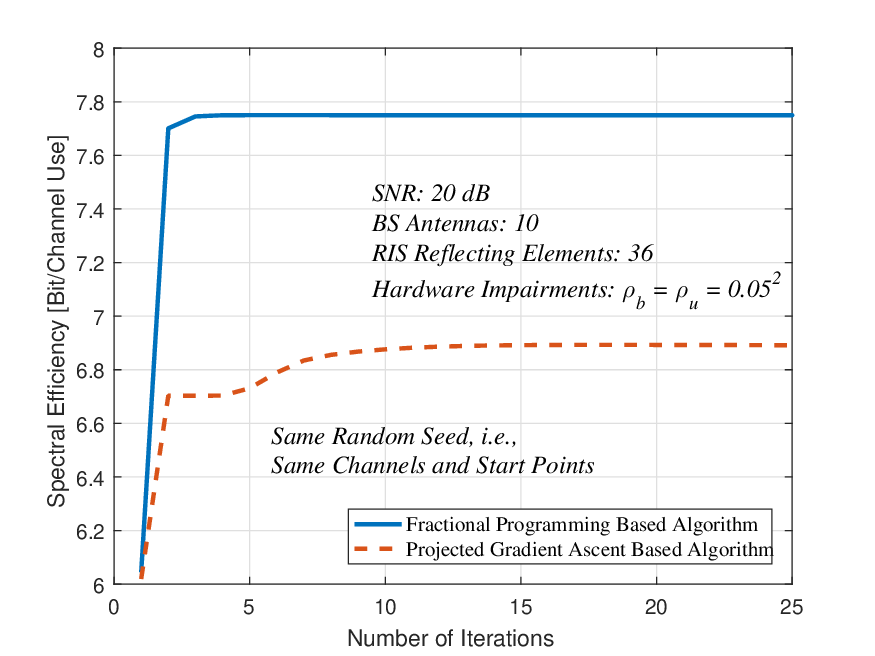}
	\caption{Convergence performance of our proposed algorithm.}
	\label{fig_1}
\end{figure}

\begin{figure}[h]
	\centering
	\includegraphics[width=3.2 in]{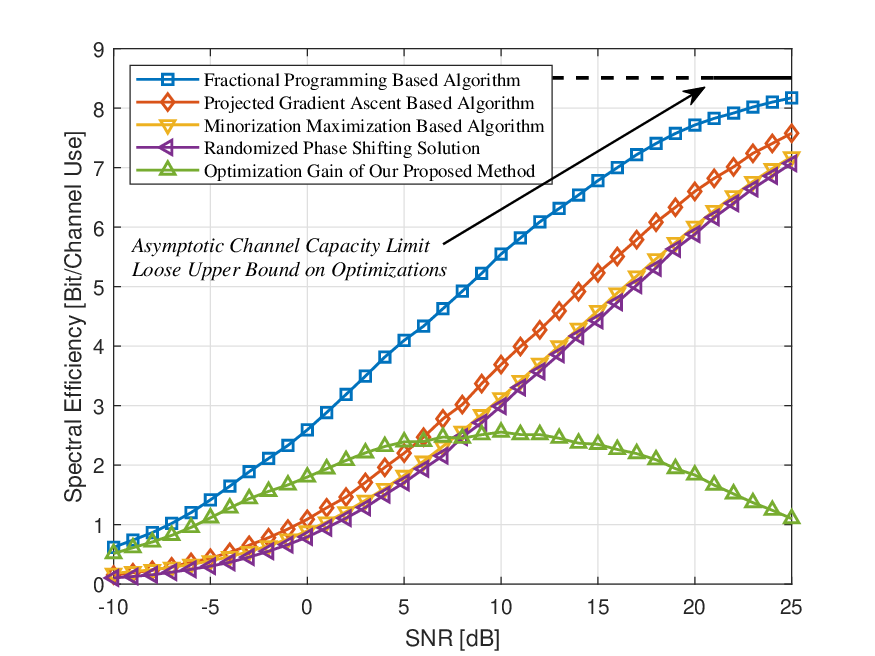}
	\caption{Optimization results of different algorithms versus SNR.}
	\label{fig_2}
\end{figure}

\begin{figure}[h]
	\vspace{- 10 pt}
	\centering
	\includegraphics[width=3.2 in]{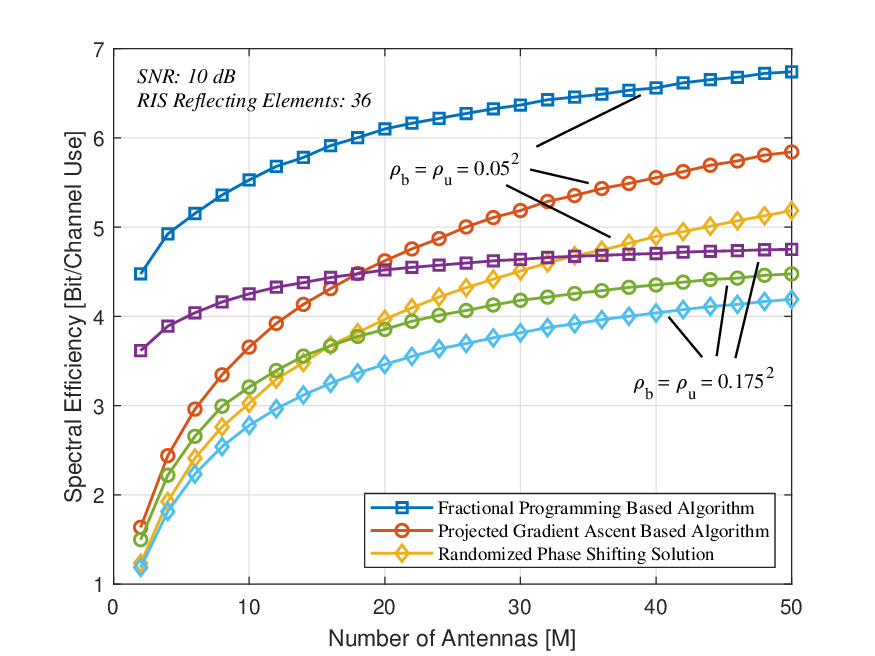}
	\caption{Optimization results of different algorithms versus the number of antennas at the BS.}
	\label{fig_3}
	\vspace{- 5 pt}
\end{figure}

We first show the convergences of our proposed algorithm and the projected gradient ascent algorithm.
In both methods, the simulation settings and start points are identical. 
It can be observed from Fig. \ref{fig_1} that
our proposed method has a fast convergence and has higher optimized value.

In Figs. \ref{fig_2} and  \ref{fig_3}, we plot the results of different algorithms versus SNR and the number of antennas, respectively, and the results are obtained through $1, 000$ Monte Carlo trials.
Because the curves of optimized results versus the number of reflecting elements are similar to that of optimized results versus SNR, we omit that here for compact pages.
In Fig. \ref{fig_2}, the number of antennas at the BS is set as $10$, and the number of reflecting elements at the RIS is set as $36$. The received SNR's at the user are selected from the set of $[ -10 \text{ dB},  25 \text{ dB}]$. The parameters of hardware impairments are set as $\rho_{\mathrm{b}} = \rho_{\mathrm{u}} = 0.05^2$.
From this figure, it is observed that our proposed algorithm has a marked enhancement in the performance, particularly when SNR is not high, within the range of $[ -15 \text{ dB},  15 \text{ dB}]$.
The performance of the MM-based algorithm \cite{9239335} is not good here since the update rule derived in \cite{9239335} may not be suitable for the case of practical phase shifts.
In Fig. \ref{fig_3}, the SNR is set as $10$ dB. This figure depicts the influence of varying the number of BS antennas on the optimization results with two different parameters of hardware impairments, i.e., $\rho_{\mathrm{b}} = \rho_{\mathrm{u}} = 0.05^2$ and $\rho_{\mathrm{b}} = \rho_{\mathrm{u}} = 0.175^2$.
It is evident from the results that our proposed algorithm is much better and even when the number of BS antennas is small, our algorithm continues to exhibit excellent performance.
Our proposed algorithm demonstrates the capability of more rapidly reaching the performance limit caused by hardware impairments.

\vspace{5 pt}
\section{Conclusion}

This paper considers both transceiver hardware impairments and practical phase shifts of RIS, and re-optimizes the passive beamforming at this scenario.
Unlike the existing works, which only consider transceiver hardware impairments and did not leverage the structure of objective functions, this paper utilizes the quadratic transform to handle the fractional structure of the optimization problem. In addition, this paper also utilizes some other optimization techniques, such as penalty-based methods and DC programming, to further solve the sub-problems.
The performance of the proposed method is much better than the existing works, especially when the SNR is not high and the number of antennas is small.
This work holds significance for the application of RIS in real-world scenarios.

\vspace{5 pt}
\bibliographystyle{IEEEtran} 
\bibliography{reference}

\vfill

\end{document}